\documentclass[a4paper,fleqn]{cas-dc}

\usepackage[numbers]{natbib}

\def\tsc#1{\csdef{#1}{\textsc{\lowercase{#1}}\xspace}}
\tsc{WGM}
\tsc{QE}
\tsc{EP}
\tsc{PMS}
\tsc{BEC}
\tsc{DE}

\begin{document}\sloppy
\let\WriteBookmarks\relax
\def\floatpagepagefraction{1}
\def\textpagefraction{.001}
\shorttitle{Author accepted preprint - IChemE ChERD \textit{Emerging Stars Special Issue} - 2021}
\shortauthors{D. T. P\'{e}rez-\'{A}lvarez et~al.}

\title [mode = title]{Foam Flows in Turbulent Liquid Exfoliation of Layered Materials and Implications for Graphene Production and Inline Characterisation}

\author[]{\textcolor{black}{\hspace{-0.2cm} Diego Tomohisa P\'{e}rez-\'{A}lvarez}}

\address[]{School of Engineering, University of Birmingham, B15 2TT, United Kingdom}

\author[]{\textcolor{black}{\hspace{-0.2cm} Philip Davies}}

\author[]{\textcolor{black}{\hspace{-0.2cm} Jason Stafford}}
\cormark[1]
\ead{j.stafford@bham.ac.uk}

\cortext[cor1]{Corresponding author}

\begin{abstract}
Surfactants are often used to stabilise two-dimensional (2D) materials in environmentally friendly solvents such as water. Aqueous-surfactant solutions prevent agglomeration of nanosheets through steric and electrostatic repulsion, facilitating the production of high concentration nanomaterial dispersions. Turbulent, shear-assisted liquid exfoliation of layered precursor materials produces defect-free nanosheets by promoting mixing and generating sufficiently high shear rates to overcome out-of-plane van der Waals bonds. In the presence of a liquid-gas interface, a consequence of using surfactants in turbulent flows is the formation of foam. In this experimental study, batch exfoliation of graphite particles into few-layer graphene was performed using a kitchen blender modified to operate across Reynolds numbers, $Re \sim 10^{5} - 10^{6}$. Foam formation during turbulent operation was found to influence the hydrodynamics of the liquid exfoliation process. Measurements on the motion of graphite particles indicate that surfactant concentration can alter the rheology of the mixture under dynamic conditions and change the material flow patterns within the device. As a result, the surfactant concentration that maximised graphene concentration was found to be non-unique. This highlights that the design and selection of surfactants should consider both molecular scale repulsion effectiveness and macroscale hydrodynamics of the liquid exfoliation process. Furthermore, the multi-phase turbulent flows and complex fluids that exist during batch exfoliation in aqueous-surfactants create major challenges for realising $\textit{in situ}$ 2D material characterisation and quality control. Here, we have developed a protocol to enable inline uv-vis-nIR spectroscopy to determine graphene production and atomic layer number changes in-process. These insights on exfoliation and characterisation of graphene in aqueous-surfactant dispersions can help advance the development of resource-efficient large-scale production of high-quality 2D materials for future technologies.

\end{abstract}

\begin{keywords}
liquid exfoliation \sep stirred tank \sep 2D materials \sep real-time monitoring
\end{keywords}

\maketitle

\section{Introduction}
\noindent The remarkable properties of graphene and related two dimensional (2D) materials have shown enormous potential to create nanotechnologies that address pressing challenges across energy, sustainability and healthcare. Although graphene is the most well-known, the library of 2D materials is extensive, including insulators such as h-BN, semi-conductors from the 2D chalcagonides family (e.g. MoS$_{2}$), and numerous 2D oxides (e.g. MoO$_{3}$) \cite{Geim2013, Nicolosi2013}. Over the past decade, laboratories across the world have used these materials to develop high-performance composites, batteries, supercapacitors, solar cells, membranes, visible light photocatalysts, and flexible sensors \cite{Ferrari2015}. 

One of the main challenges restricting the translation of these promising materials beyond laboratory settings and into real world applications is large-scale production of high quality materials \cite{Ferrari2015}. Insufficient nanomaterial production rates and quality control are key areas that require attention \cite{Nicolosi2013}. Addressing this will improve the economic viability and lower the risk for manufacturers who incorporate these materials into future technologies \cite{Stafford2018}.

Liquid exfoliation has emerged as a prime candidate for large-scale production of 2D materials. This top-down method disperses a layered precursor material (e.g. graphite) in a suitable solvent and separates the layers in solution using either mechanical, chemical or electro-chemical methods to create nanomaterial dispersions (e.g. mono- and few-layer graphene) \cite{Stafford2018}. Mechanical methods that rely predominantly on shear-driven exfoliation are beneficial as they typically produce defect-free 2D materials without oxidation. Sonication \cite{Hernandez2008}, high-shear mixing \cite{Paton2014}, microfluidisation \cite{Paton2017}, Taylor-Couette and thin film reactors \cite{Stafford2021, Chen2014} are a few of many examples that have been used successfully.

A number of these utilise well-established techniques used extensively across the chemical engineering research and design community. Batch approaches such as high-shear mixing and turbulent stirred tank reactors have been demonstrated as suitable platforms for shear exfoliation of graphene, h-BN, MoS$_2$ and WS$_2$ \cite{Paton2014,Varrla2014, Yi2014, Biccai2019}. These approaches have been studied experimentally and using computational fluid dynamics to provide insights on modes of nanomaterial breakup \cite{Zhang2012}, coherent flow structures \cite{Bakker2004,Janiga2019}, turbulent statistics \cite{Vikash2019} and mixing effectiveness \cite{Nienow1997}.           
The overlap with established batch approaches provides great opportunity to support the physical design of scaled-up solutions that produce 2D materials and nanocomposite blends on an industrial scale. However, optimising production rate, material quality and waste requires further exploration of the underlying mechanisms which are less well understood. The hydrodynamics driving exfoliation span from the tank scale down to the 2D nanosheets with atomic scale thicknesses. 

Molecular simulations and micromechanical modelling of the controlled separation of graphene sheets have provided insights on the shear and peeling mechanisms at the nanoscale \cite{Sinclair2018,Salussolia2020}. Experiments by Paton et al. \cite{Paton2014} and Biccai et al. \cite{Biccai2019} showed there is a minimum shear rate required to overcome van der Waals attractive forces and exfoliate layered materials. Stafford et al. \cite{Stafford2021} recently uncovered the general scaling relationship governing production for continuous flow, shear exfoliation systems. Combining high fidelity numerical simulations (DNS, LES), fluid dynamic experiments and nanomaterial characterisation, it was found that graphene concentration depends on derived parameters of shear rate and particle residence time. Transmission and scanning electron microscopy on individual nanosheets showed signatures of slip and peel exfoliation mechanisms at the nanoscale. In shear exfoliation environments generated by turbulent Taylor-Couette flows, surface erosion of the precursor microscale particles was observed. More aggressive break-up of graphite precursor particles was shown for cavitation-driven exfoliation using sonication \cite{Zheling2020}.

Solvent choice also influences nanomaterial concentration in liquid exfoliation processes \cite{Hernandez2010}. Many high performance solvents are toxic and have high boiling points which has led to research on greener alternatives \cite{Stafford2018,Salavagione2017}. Water can function as a suitable solvent for few-layer graphene dispersions when modified through co-solvent strategies \cite{Stafford2021}  or with the addition of surfactants \cite{Lotya2009, Lotya2010}. In the latter approach, surface-active agents adsorb to nanosheets and prevent re-stacking through steric and electrostatic repulsion mechanisms. This has been found to enable high concentration formulations in water (>1 mg/mL)\cite{Varrla2014}.

\begin{figure*}
	\centering
		\includegraphics[scale=0.68]{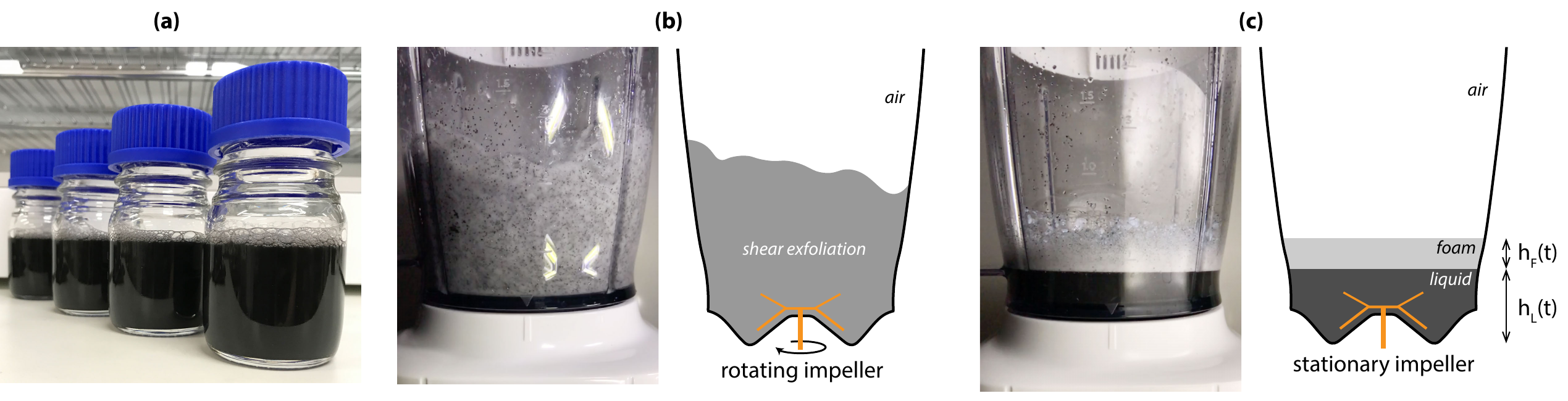}
	\caption{Shear exfoliation in aqueous-surfactant dispersions can generate foam flows. (a) Liquid dispersions of few-layer graphene produced by the batch exfoliation process. (b) Foamy fluid mixture during turbulent shear exfoliation in a kitchen blender. (c) Gravitational phase separation when the process stops.}
	\label{fig:1}
\end{figure*}

These previous studies suggest that the synthesis route from precursor to nanosheet can be sensitive to the approach. They also suggest that the predominant impact of adding surface-active agents in aqueous-surfactant dispersions of 2D materials is molecular scale repulsion at the nanosheet scale. In this study, we investigate batch exfoliation using a kitchen blender to produce graphene in aqueous-surfactant dispersions and examine secondary foam formation effects that are currently neglected. This system is capable of exfoliating 2D materials \cite{Varrla2014,Yi2014} and can serve as a template for studying contemporary stirred tank reactors that are used at larger scales for production and material blending operations. A reliance on slow and expensive \textit{ex situ} materials characterisation tools also presents a bottleneck for realising large-scale high-throughput material quality control \cite{Backes2016,Stafford2021}. Here, turbulent mixing and shear exfoliation have been investigated using inline spectroscopy to probe the process characteristics \textit{in situ} and exfoliation performance on-the-fly.     

\begin{figure*}
	\centering
		\includegraphics[scale=0.85]{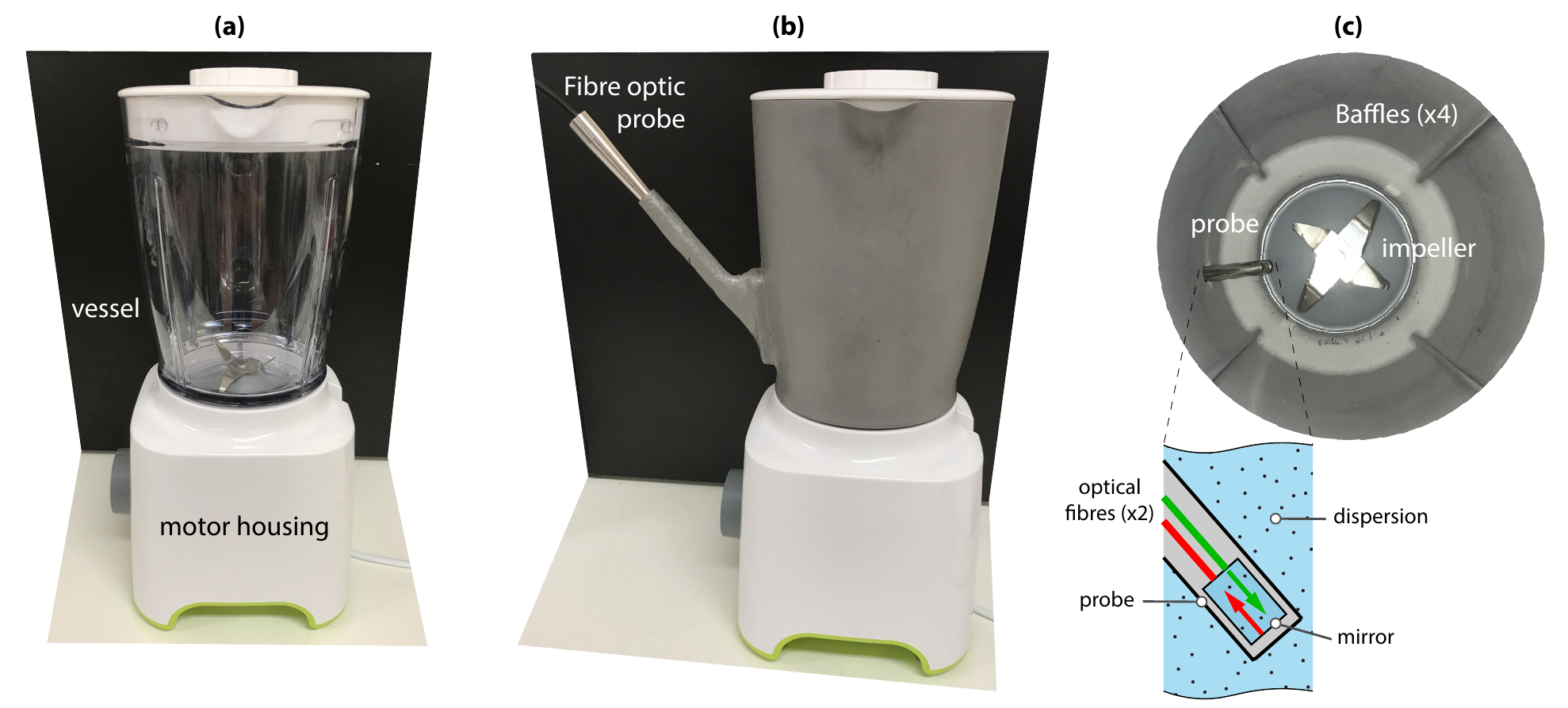}
	\caption{Implementation of inline uv-vis-nIR spectroscopy for production monitoring. (a) The standard kitchen blender vessel. (b) The modified vessel with fibre optic probe insertion. (b) Internal view of the vessel showing the location of the probe for \textit{in situ} spectroscopy.}
	\label{fig:2}
\end{figure*}

\begin{figure*}
	\centering
		\includegraphics[scale=0.95]{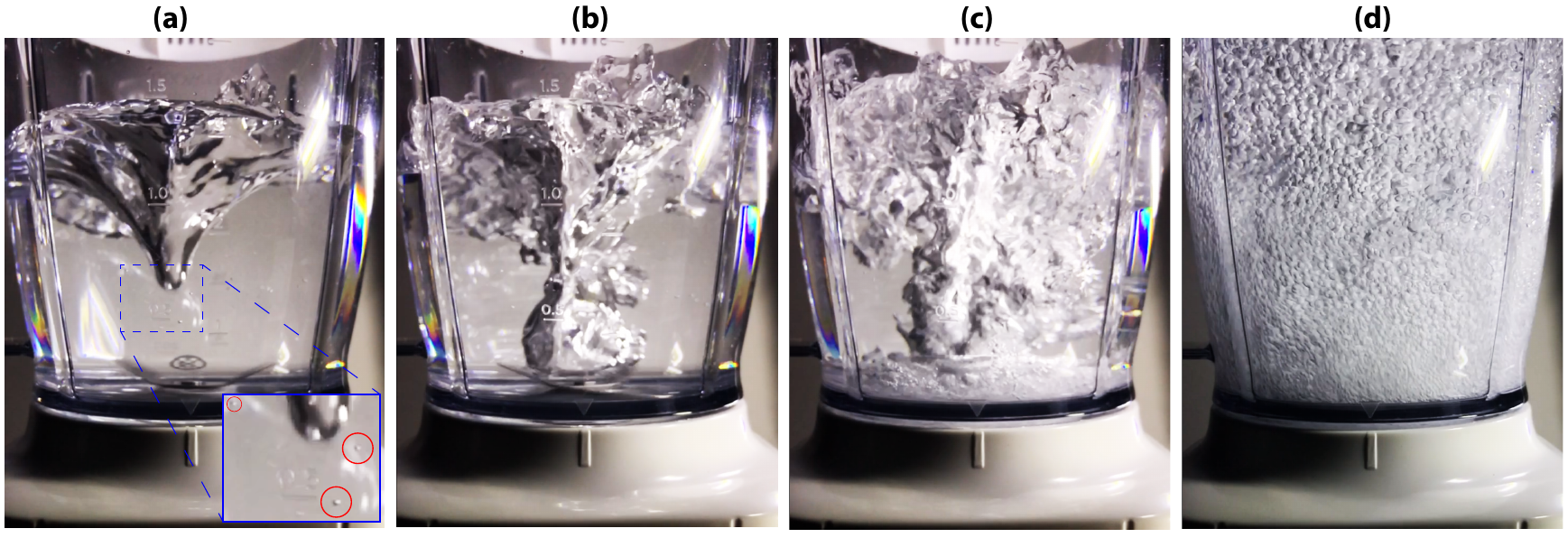}
	\caption{Evolution of the liquid-gas behaviour upon impeller start-up. (a) $\omega{t} = 2100$. (b) $\omega{t} = 2160$. (c) $\omega{t} = 2210$. (d) $\omega{t} > 7330$. Images shown are for a liquid volume, $V = 800$ mL, $Re \approx 2.6 \times 10^{6}$, $Fr \approx 6.1 \times 10^3$, and $We \approx 2.5 \times 10^6$.}
	\label{fig:3}
\end{figure*}

\section{Materials and Methods}

\subsection{Experimental arrangement}
\noindent Batch production of few-layer graphene using turbulent shear-assisted liquid exfoliation was performed using a 350W kitchen blender (Figs. \ref{fig:1} and \ref{fig:2}, Kenwood BLP31). This was used as a model system for traditional stirred mixers which operate across a broad range of scales from lab to industry. The kitchen blender has a contour vessel shape and a 4-blade impeller design with two blades angled upwards and two facing downwards. This configuration promotes solids suspension \cite{Harnby1992}. The diameter of these upper and lower blades differs ($D_{u} = 48$ mm; $D_{l} = 62$ mm) and an average diameter of $D = 55$ mm has been used for the calculation of non-dimensional parameters. The vessel contains four baffles and is tapered with an average diameter $\approx 110$ mm. These baffles are 5 mm in width and extend beyond the full height of all liquid volumes investigated (200 mL $\leqslant V \leqslant$ 800 mL). 

The standard kitchen blender was restricted to two speed settings. The motor control circuit was modified and a 10A AC phase angle power regulator was installed (United Automation, QVR-TB-RFI). This allowed variable speed control from $500\pm{100}$ rpm $<\omega<$ $20000\pm{1500}$ rpm and the impeller speed was monitored during operation using an optical tachometer (Omega HHT13). The resulting non-dimensional numbers based on the properties of water were: Reynolds number, $Re = \rho \omega D^2/\mu \sim 10^5 - 10^6$, Froude number, $Fr = \omega^2 D / g \sim 10^{1} - 10^{4}$, and Weber number, $We = \rho \omega^2 D^3 / \sigma \sim 10^3 - 10^7$. This indicates that fluid mixing in a vessel containing water only is dominated by the applied inertial forces. The confirmation of this is shown in the flow observations presented in Fig. \ref{fig:3} for an intermediate impeller speed and $Re \approx 2.6 \times 10^{6}$. 

Graphite flakes (Sigma Aldrich, 332461) were dispersed in de-ionised water (5 M$\Omega$m) and a surfactant (Fairy Liquid). The surfactant is a household detergent that has been shown to be an effective stabilising agent for graphene in water \cite{Varrla2014}. This detergent contains 15-30\% anionic surfactants and 5-15\% non-ionic surfactants. During shear exfoliation the fluid is a mixture of unexfoliated graphite particles, exfoliated few-layer graphene, foam and liquid (Fig. \ref{fig:1}b). When the process is stopped, gravity separates out the foam and liquid phases into two distinct layers (Fig. \ref{fig:1}c). The height of these layers is time-dependent with liquid draining to the base of the vessel from the thin films and Plateau borders within the foam structure \cite{Kraynik1988}. 

The foam formation process can be explained in part by examining the evolution of the liquid-gas interface for the case of water without surfactant. Figure \ref{fig:3} shows snapshots during the early stages of impeller start-up and to the point where quasi-steady bubbly flow exists throughout. The vessel has a large diameter to baffle width ratio $\approx 22$ that typically suppresses vortexing for fluids with high viscosity \cite{Holland1966}. Water is a low viscosity liquid and a vortex forms in this system leading to a drop in liquid level in the core (Fig. \ref{fig:3}a). A small number of cavitation microbubbles were also observed at this Reynolds number $Re \approx 2.6 \times 10^6$. Vortexing pulls air into the path of the rapidly rotating impeller and when this interacts with the blades it produces large quantities of small gas bubbles that are forced outwards (Fig. \ref{fig:3}b,c). These propagate throughout the vessel and the resulting bubbly flow pattern persists indefinitely (Fig. \ref{fig:3}d). Smaller bubbles near the base of the vessel in the impeller region rise and coalesce to form larger bubbles near the top. In the presence of surfactants, this aeration process leads to foam formation as surfactant molecules concentrate at liquid-gas interfaces preventing bubble coalescence and lowering surface tension. 

\subsection{Material synthesis and \textit{ex situ} characterisation}
\begin{figure}
	\centering
		\includegraphics[scale=0.6]{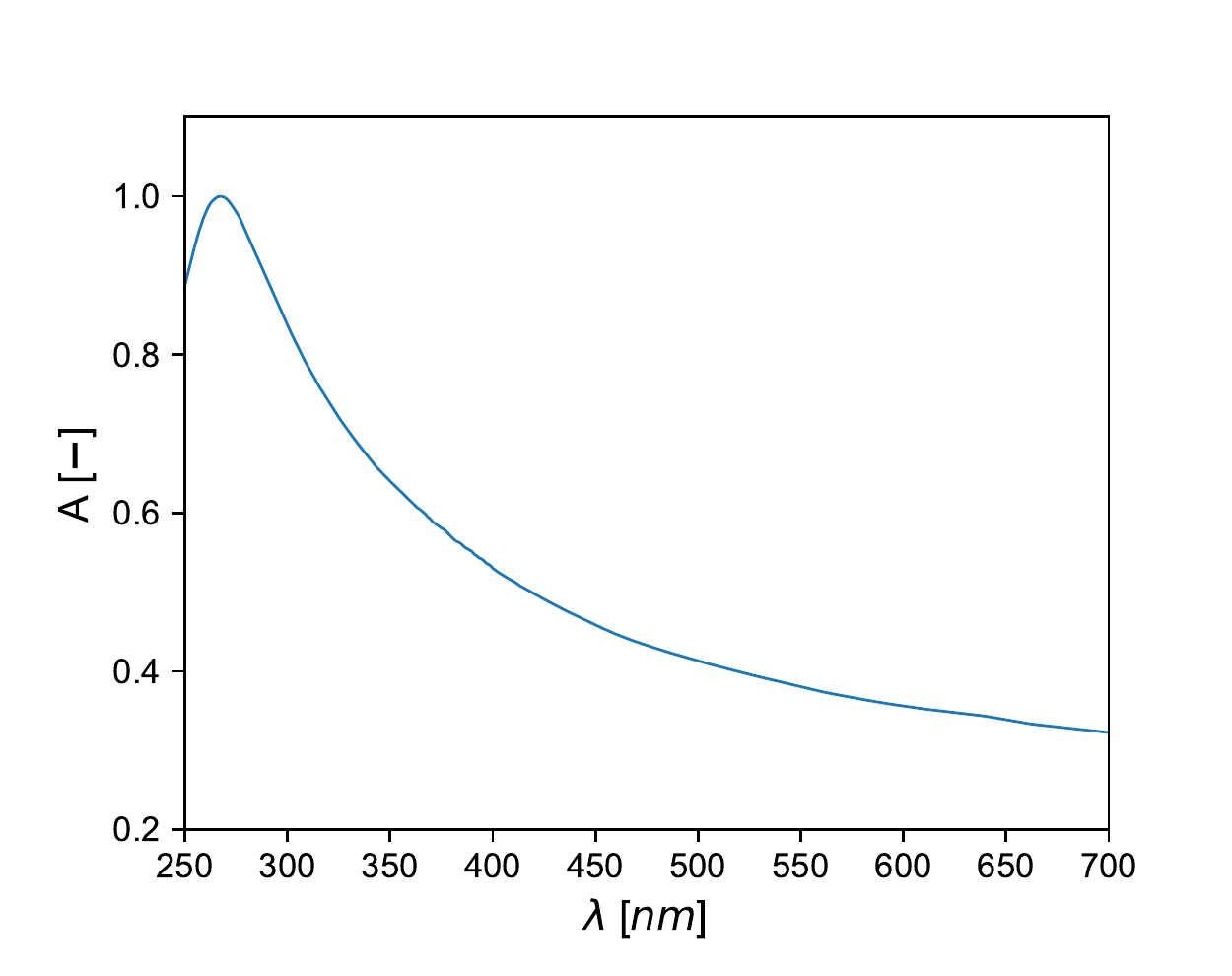}
	\caption{Extinction spectrum of few-layer graphene. The peak visible is situated at $\lambda=267$ nm.}
	\label{fig:4}
\end{figure}

\noindent Aqueous-surfactant mixtures of graphite particles were processed over time using a 1 minute ON followed by 1 minute OFF operating procedure \cite{Varrla2014}. This was necessary as the manufacturer's recommend a maximum 1 minute operating interval to avoid overheating the motor. Results have been presented using a processing time that corresponds to the impeller ON time (e.g. a 15 minute process time required operating the blender 1 min ON / 1 min OFF for 30 minutes). Few-layer graphene (FLG) were separated from the mixture by post-process centrifugation. At the end of each test, 15 mL samples were pipetted into centrifugation tubes from the vessel. Samples were cetrifuged at 1500 rpm (214 RCF) for 45 mins and the top 5 mL supernatant containing few-layer graphene (FLG) was removed for analysis. Confirmation of FLG with average atomic layer number $N_{FLG} < 10$ was obtained by measuring the extinction spectra and applying spectroscopic metrics to calculate average layer number using the relationship $N_{FLG} = 25(A_{550}/A_{max}) - 4.2$ \cite{Backes2016}. An example of a FLG extinction spectrum for 250 nm $ < \lambda < 700$ nm is shown in Figure \ref{fig:4}. Concentration ($C_g$) was measured by performing uv-vis-nIR spectroscopy on samples at a wavelength where absorbance is unaffected by nanomaterial size and thickness. This was quantified using extinction measurements at $\lambda = 660$ nm, an extinction coefficient of $\varepsilon_{660} = 6600$ mL mg$^{-1}$ m$^{-1}$ \cite{Varrla2014}, and the Lambert-Beer relationship, $C_g = A_{660}/\varepsilon_{660} L$, where $L = 10$ mm was the optical path length of the quartz cuvette containing the sample.        

\subsection{\textit{In situ} materials characterisation} \label{Insitu-setup}
\noindent Investigations were performed to assess if inline spectroscopy could be implemented in turbulent batch exfoliation of aqueous-surfactant dispersions of 2D materials. The vessel was modified to allow insertion of a stainless steel probe into the process volume (Fig. \ref{fig:2}). This probe housed two fused silica optical fibres. One fibre transmitted light from the spectrophotometer source into the measurement region. This light travelled through the fluid medium, reflected off a mirror, and returned to the spectrophotometer through the second fibre. The probe was mounted at an obtuse angle and located near the base of the vessel to 1) lower the risk of solids depositing on the mirror surface and 2) ensure the probe was submerged for all process volumes examined in this study. The vessel was painted opaque to block light from external sources in the laboratory and avoid interference with the fibre optic probe.    

\begin{figure*}
    \hspace{0cm}\textbf{(a)}\hspace{8cm}\textbf{(b)}\hspace{7cm}
    \includegraphics[scale=0.65]{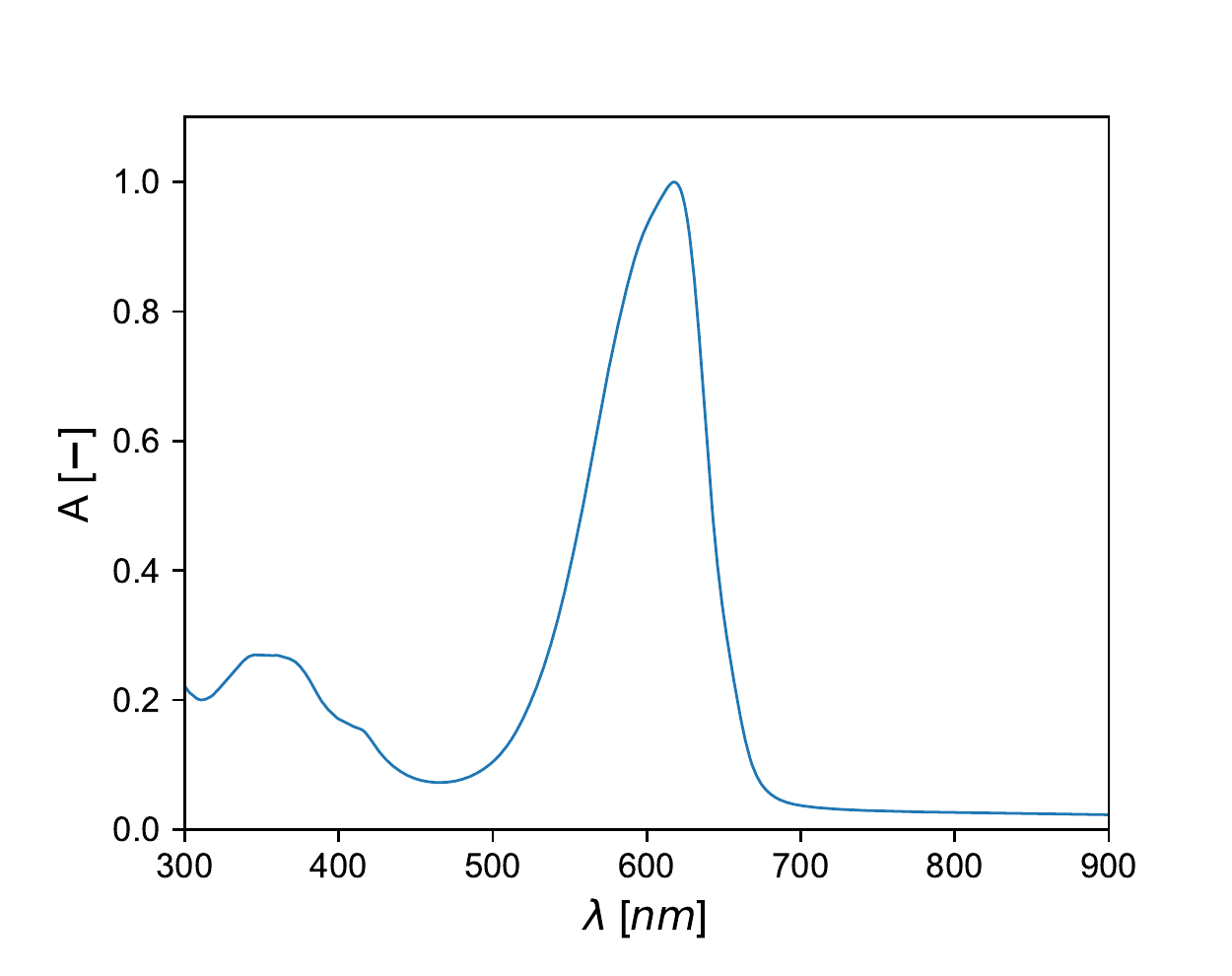}
    \centering
    \includegraphics[scale=0.65]{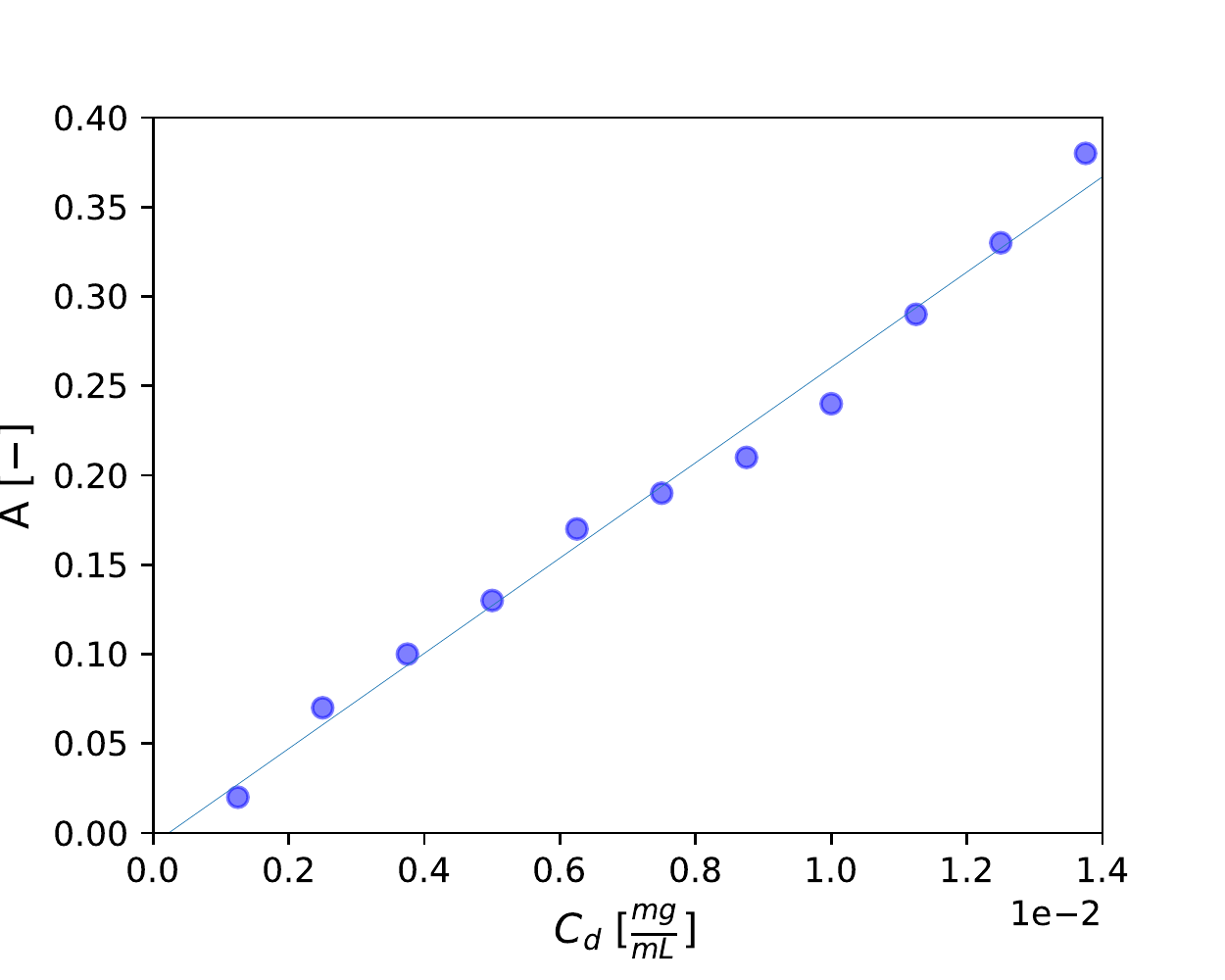}
    \caption{(a) Absorbance spectrum of blue food dye. The peak visible is situated at $\lambda=617$ nm. (b) Relationship between the concentration of blue food dye ($C_d$) and absorbance measured using \textit{in situ} spectroscopy at $\lambda=617$ nm and $V = 400$ mL.}
    \label{fig:5}
\end{figure*}

Initial trials were performed using a simpler solution without surfactant or graphite. Blue food dye (Spirulina, Glycerine, Water) was added to a vessel filled with 400 mL of water. The blue dye was found to have an absorbance peak at $\lambda = 617$ nm using \textit{ex situ} uv-vis-nIR spectroscopy (Fig. \ref{fig:5}a). This wavelength was subsequently chosen to measure changes in absorbance due to dye concentration changes. During mixing, the formation and transport of gas bubbles disturbs the optical measurements making it difficult to detect dye concentration changes. During the OFF period however, gas bubbles coalesced, absorbance readings stabilised and were solely dependent on the dye concentration in solution. Figure \ref{fig:5}b shows these changes in dye concentration together with a linear fit that follows the anticipated Lambert-Beer relationship, $A = \varepsilon C L$,  where the slope is $\varepsilon_{617} \approx 2600$ mL mg$^{-1}$ m$^{-1}$. These trials confirm that the inline measurement system is capable of detecting concentration changes when considerations are made for the additional gas phase that is present during operation. In the following results sections, the inline characterisation method has been extended to the batch exfoliation process where foam flows and graphitic materials exist. 

\subsection{Flow characterisation} \label{MixMethod}
\noindent The flow behaviour during turbulent exfoliation inside the kitchen blender has been investigated to explore connections between the hydrodynamics of the multiphase process and the nanomaterial produced. Using a clear vessel illuminated with a $4\times2$W LED light source, high speed images were acquired at 720p and 240 fps for an intermediate impeller speed, $\omega = 10^4$ rpm. This was also used to investigate the impeller startup process and bubbly flows discussed previously (Fig. \ref{fig:3}). 

\begin{figure*}
	\centering
		\includegraphics[scale=1.07]{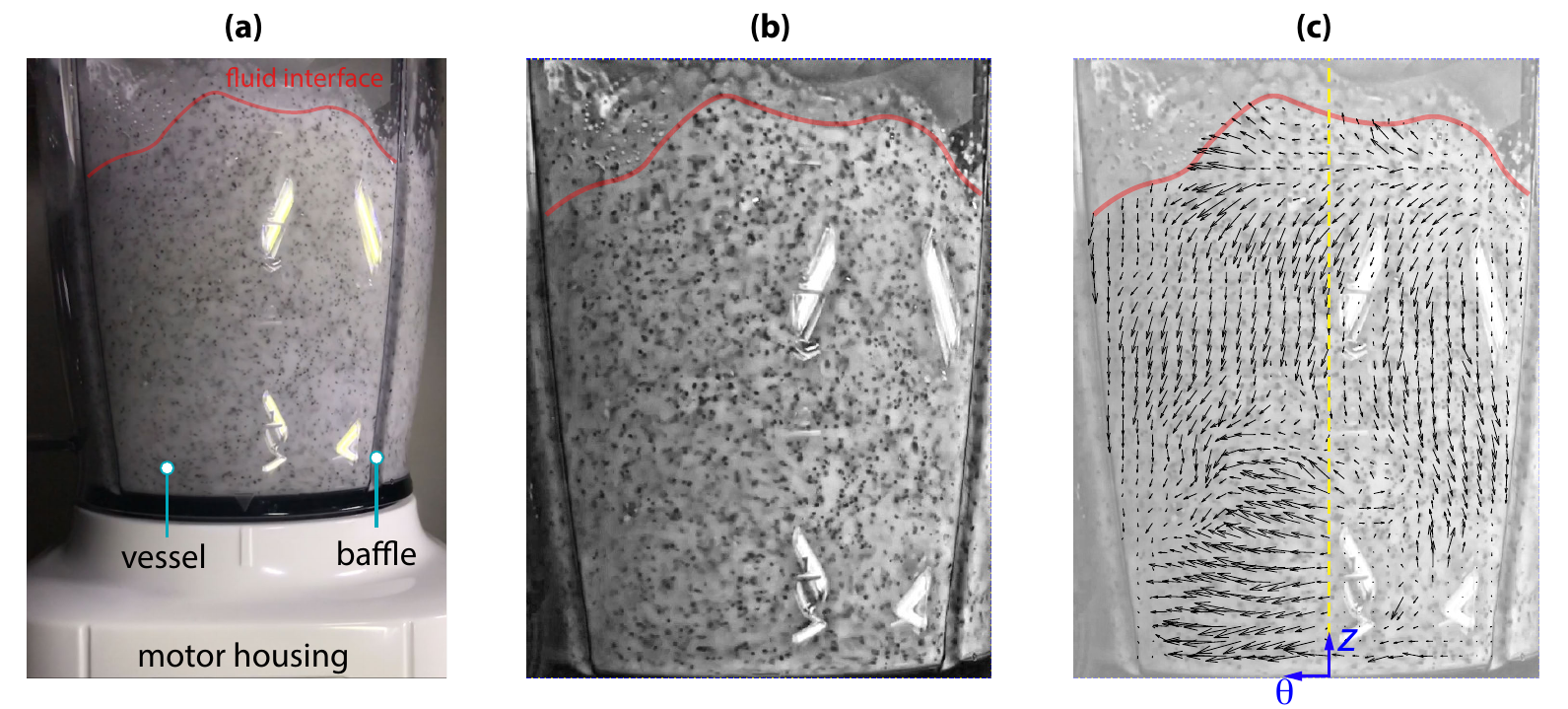}
	\caption{Near-wall graphite particle image velocimetry. (a) Raw high-speed image captured during liquid exfoliation. (b) contrast-enhanced region of interest. (c) instantaneous displacement vector field as viewed from the camera perspective. Images shown are for $V = 200$ mL, $\omega = 10^{4}$ rpm, initial graphite concentration, $C_i = 20$ mg/mL and surfactant concentration, $C_i/C_{fl} = 8$.}
	\label{fig:6}
\end{figure*}

Images were acquired during shear exfoliation in aqueous-surfactant dispersions to examine the graphite material flows within the foam mixtures. An example of one of these images is shown in Figure \ref{fig:6}a. Dark graphite particles with an average diameter $\approx$500 $\mathrm{\mu m}$ are visible within the bright foam structure (Fig. \ref{fig:6}b). This contrast was exploited to perform particle image velocimetry (PIV) and reveal the graphite flow behaviour in the near-wall region of the vessel (Fig. \ref{fig:6}c). 

Recordings began after the impeller start-up phase and for $\omega{t} > 10^4$, when the aqueous-surfactant dispersion was in a quasi-steady state. A total of 3000 image pairs were recorded and analysed to determine graphite particle displacements for each image pair (Fig. \ref{fig:6}c) \cite{PIVlab}. Notably, these particle displacements were from the perspective of the camera viewpoint. The vessel has a circular cross-section that varies along its height and the captured particle motions are in three-dimensions. As a result, velocity magnitudes have not been resolved. For the purpose of this study, quantitative measurements on the particle flow direction were obtained and presented in the perspective view. Ensemble-averaged streamlines were calculated from the instantaneous graphite displacement fields in the near-wall region. The ensemble average was based on a sample size of 3000 to capture statistically significant information in turbulent flows \cite{Stafford2012}. This ensured the measurements accurately represented the time-averaged graphite flow behaviour for the entire liquid exfoliation process.

\section{Results and discussion}

\subsection{Liquid exfoliation of graphene}
\begin{figure}[t]
	\centering
		\includegraphics[scale=0.65]{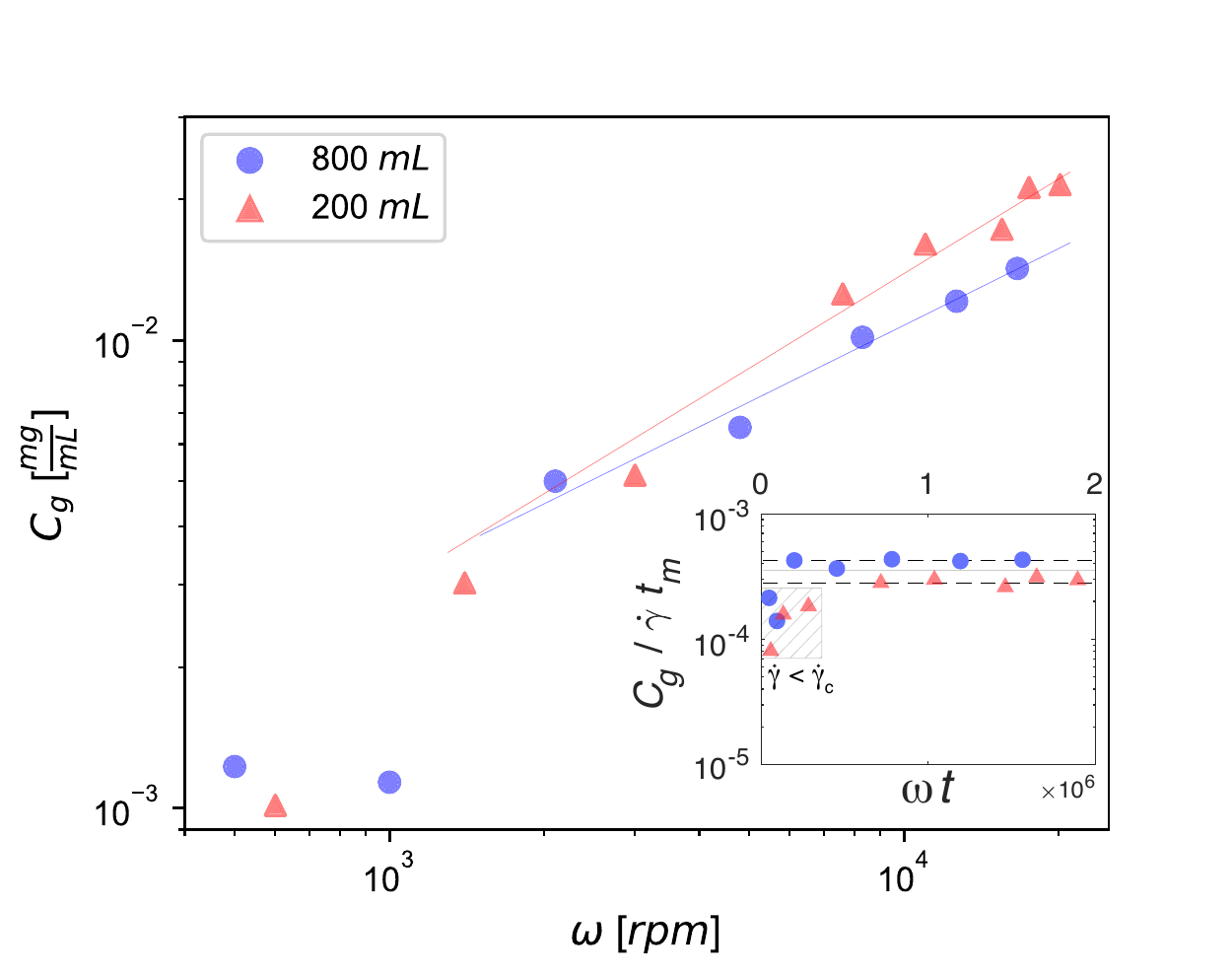}
	\caption{Concentration of graphene for different impeller rotational speeds ($\omega$) and fluid volumes ($V$). All concentration data were measured at a process time, $t = 15$ mins.}
	\label{fig:7}
\end{figure}

\noindent The concentration of graphene produced after 15 mins processing in the kitchen blender for different impeller speeds and fluid volumes is shown in Figure \ref{fig:7}. At speeds higher than $\omega \approx 1500$ rpm, the concentration increases from $C_g \sim 10^{-3}$ mg/mL to $C_g \sim 10^{-2}$ mg/mL, with the smaller 200 mL volume process producing higher concentrations of few-layer graphene.

The kitchen blender is similar to a stirred tank reactor with a power input per unit volume that is related to the shear rate within the fluid, $P/V = \mu\dot{\gamma}^{2}$ \cite{SanchezPerez2006}. The power input can be expressed through the non-dimensional power number, $Po = P/\rho\omega^3D^5$, which is sensitive to flow regime. In the turbulent regime ($Re > 10^4$), $Po\approx$ constant for baffled vessels, or a slightly negative slope for unbaffled vessels \cite{Holland1966}. Assuming a constant here for simplicity ($Po = C_1$), the average shear rate becomes:

\begin{equation}
    \dot{\gamma} = \left(\frac{C_1 \rho \omega^3 D^5}{\mu V}\right)^{1/2}\label{eq:1}
\end{equation}

\noindent For the blender parameters directly varied in this work, this relationship suggests the scaling, $\dot{\gamma} \sim \omega^{3/2}V^{-1/2}$. The increase in shear rate with decrease in volume explains the higher graphene concentration measured for the smaller volume in Figure \ref{fig:7}. 

A notable reduction in graphene concentration is observed in Figure \ref{fig:7} at rotational speeds, $\omega < 1000$ rpm, similar to the findings by Varrla et al. \cite{Varrla2014}. At lower impeller speeds of $\omega \sim 10^2$ rpm, the shear rate is $\dot{\gamma} \sim 10^{1}-10^{2}$ s$^{-1}$, far below the critical criterion for overcoming the van der Waals attractive force that holds graphene layers together ($\dot{\gamma}_c \approx 10^4$ s$^{-1}$) \cite{Paton2014}. In the speed range $\omega \sim 10^3 - 10^4$ rpm, the shear rate increases to $\dot{\gamma} \sim 10^{3}-10^{5}$ s$^{-1}$ and graphene nanosheets are separated from the graphite precursor particles.

\begin{figure}[t]
	\centering
		\includegraphics[scale=0.65]{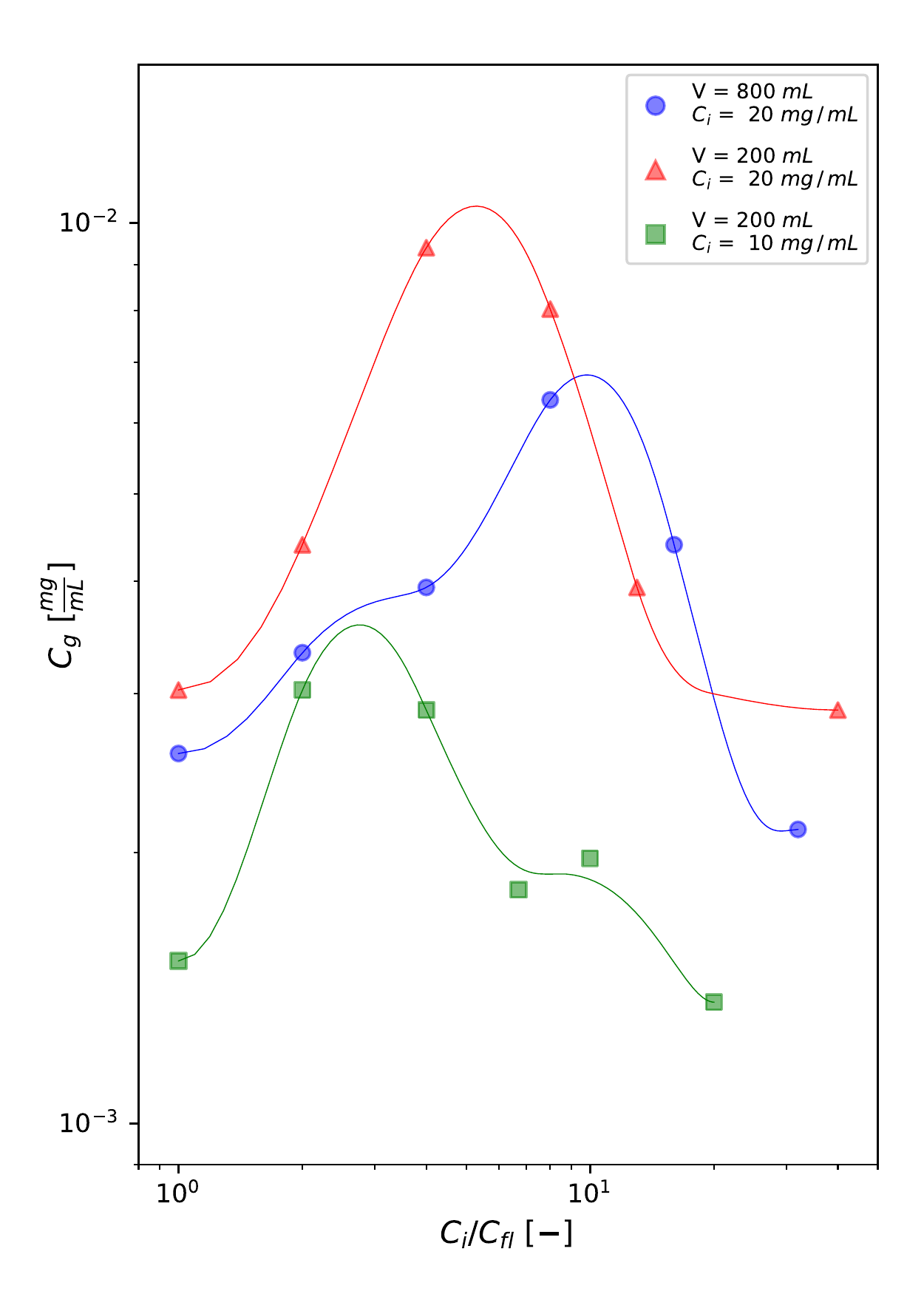}
	\caption{Concentration of graphene for different surfactant concentrations ($C_{fl}$), fluid volumes ($V$) and starting graphite concentrations ($C_i$). All graphene concentration data were measured at a process time, $t = 15$ mins, and a constant rotational speed, $\omega = 10000$ rpm..}
	\label{fig:8}
\end{figure}

In addition to strain rate, precursor residence time ($t_{res}$) has been shown to play an equally important role in 2D material production by shear exfoliation in continuous flow laminar and turbulent flow regimes. Stafford et al. \cite{Stafford2021} showed for shear exfoliation using turbulent Taylor-Couette flows that $C_g \sim \dot{\gamma} t_{res}$ when operating above the critical shear rate for exfoliation. In the high Reynolds number flows examined here, the macroscale mixing time is dominated by convective fluid motions due to the mean flow and turbulent diffusion \cite{Nagata1975}. The turbulent Peclet number, $Pe = Re\cdot{Sc}$, takes a constant value and the non-dimensional mixing time is constant irrespective of vessel volume, $ t^* = \omega t_m$. Making the assumption that fast mixing times correspond with short graphite particle residence times in the high shear regions of the vessel, this characteristic exfoliation time scales as $t_m \sim t_{res} \sim \omega^{-1}$. Concentration has been rescaled with $\dot{\gamma} t_{m}$ on the inset plot in Figure \ref{fig:7} with the collapsed data showing a reasonable agreement ($\approx{20}\%$) considering the simplified assumptions of isotropic turbulence and Newtonian fluid in eq. (\ref{eq:1}). A departure from this scaling relation occurs when $\dot{\gamma} < \dot{\gamma}_c$. This indicates that the important derived hydrodynamic parameters in turbulent batch exfoliation of 2D materials are also strain rate and mixing time.    

\subsection{Variation in surfactant concentration}
\noindent The influence of surfactant concentration on graphene production is shown in Figure \ref{fig:8}. For each process volume and initial starting graphite concentration examined, the concentration of graphene reaches a maximum value. The occurrence of a peak in graphene concentration agrees with previous studies on aqueous-surfactant dispersions using sonication with sodium dodecylbenzene sulfonate \cite{Lotya2009}, sonication with sodium cholate \cite{Lotya2010}, and turbulent shear exfoliation with detergent \cite{Varrla2014}. This optimum surfactant concentration was initially thought to be located close to the critical micelle concentration (CMC). However, it has been found to occur at much lower concentrations than the CMC and more recently suggested to be dependent on the starting graphite concentration, $C_i$. 

For $V = 800$ mL and $C_i = 20$ mg/mL, the optimum concentration $C_i/C_{fl} \approx 10$ agrees closely with the work of Varrla et al. \cite{Varrla2014} who showed for the same initial graphite concentration and a lower process volume, $V = 500$ mL, that $C_i/C_{fl} \approx 8$. At a process volume $V = 200$ mL, a further departure from this value was observed and the optimum surfactant concentration decreased to $C_i/C_{fl} \approx 3-4$. This trend suggests that the liquid exfoliation process has an important role in addition to the molecular scale nanosheet-surfactant coverage mechanisms that have been the focus of previous studies. Here, the turbulent liquid exfoliation process is sensitive to volume and this impacts the final surfactant concentration that produces the highest graphene concentrations.  

\subsection{Kitchen blender hydrodynamics}\label{hydro}

\begin{figure*}
	\centering
		\includegraphics[scale=0.95]{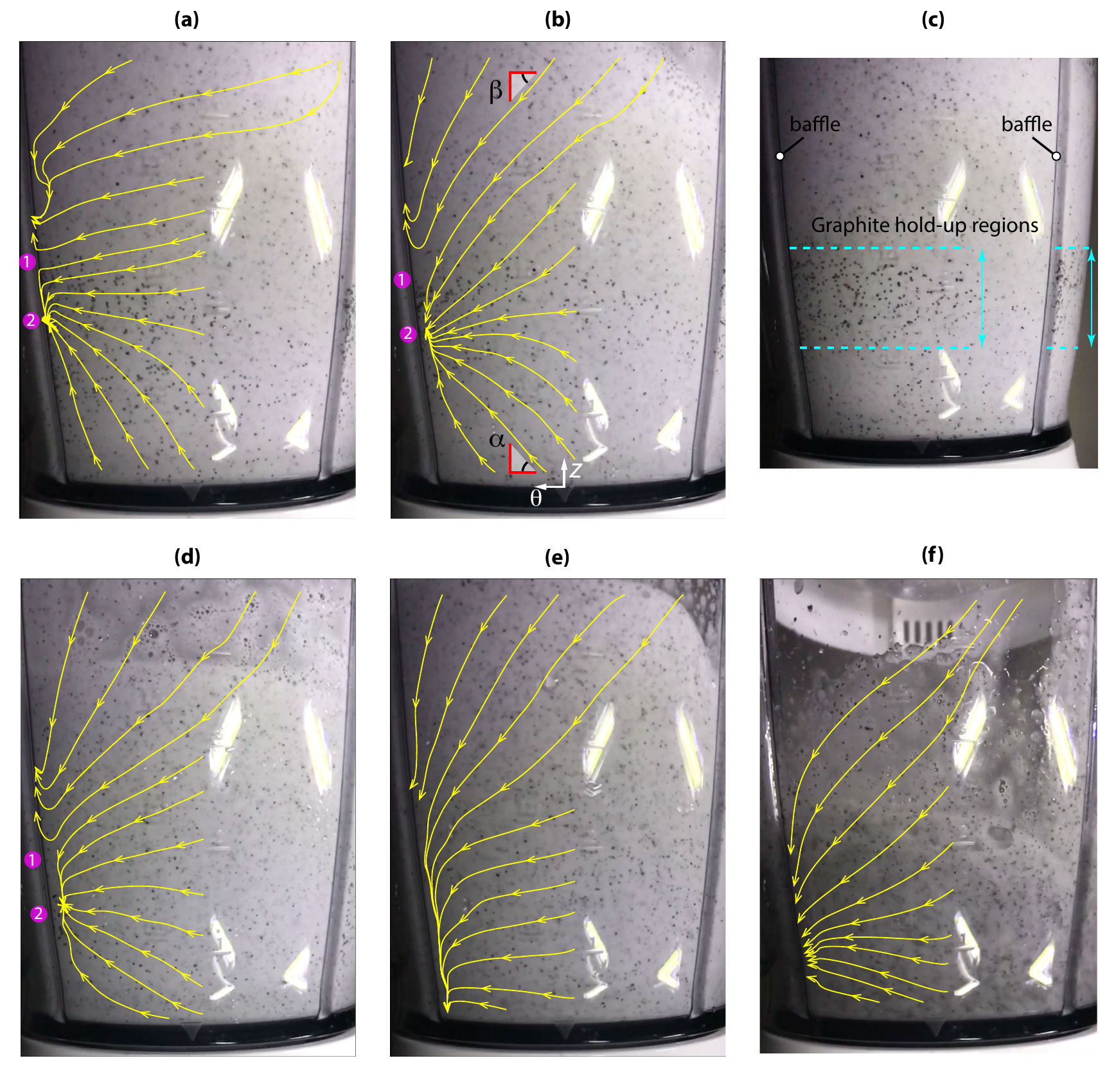}
	\caption{Ensemble-averaged near-wall graphite particle streamlines. (a) $C_i/C_{fl} = 1$. (b) $C_i/C_{fl} = 2$. (c) Graphite particle hold-up regions for $C_i/C_{fl} = 1$. (d) $C_i/C_{fl} = 4$. (e) $C_i/C_{fl} = 8$. (f) $C_i/C_{fl} = 40$. Streamlines have been ignored in the regions with light reflections. All data shown for $V = 200$ mL, $C_i = 20$ g/L, $\omega = 10000$ rpm.}
	\label{fig:9}
\end{figure*}

\noindent Although the shear rate is a function of the volume (eq. \ref{eq:1}), this alone is insufficient to explain the shift in optimum surfactant choice in Figure \ref{fig:8}. However, different volumes and surfactant concentrations were found to influence foam formation, structure, and stability. This, in turn, changed the fluid rheology under turbulent shear exfoliation conditions. Figure \ref{fig:9} shows the time-averaged near-wall graphite flow behaviour for the lowest volume and different surfactant concentrations investigated in this work. In the presence of surfactant, the aeration process described in Figure \ref{fig:3} leads to the formation of stable foams containing fine gas bubbles. The rotation of the impeller within the contoured base spreads the foam and graphite dispersion outwards from the centre of the vessel. This is where the highest particle velocities exist and is captured by the streamlines in the lower half of the vessel. The surfactant concentration varies the axial-circumferential components of velocity in this region ($\bar{u}_z, \bar{u}_{\theta}$), altering the flow angle, $\alpha$. This upward flow angle reduces from $\alpha = 60^{\circ}$ for the highest surfactant concentration $C_i/C_{fl} = 1$, to $\alpha = 17^{\circ}$ for the lowest $C_i/C_{fl} = 40$. 

\begin{figure}
	\centering
		\includegraphics[scale=0.6]{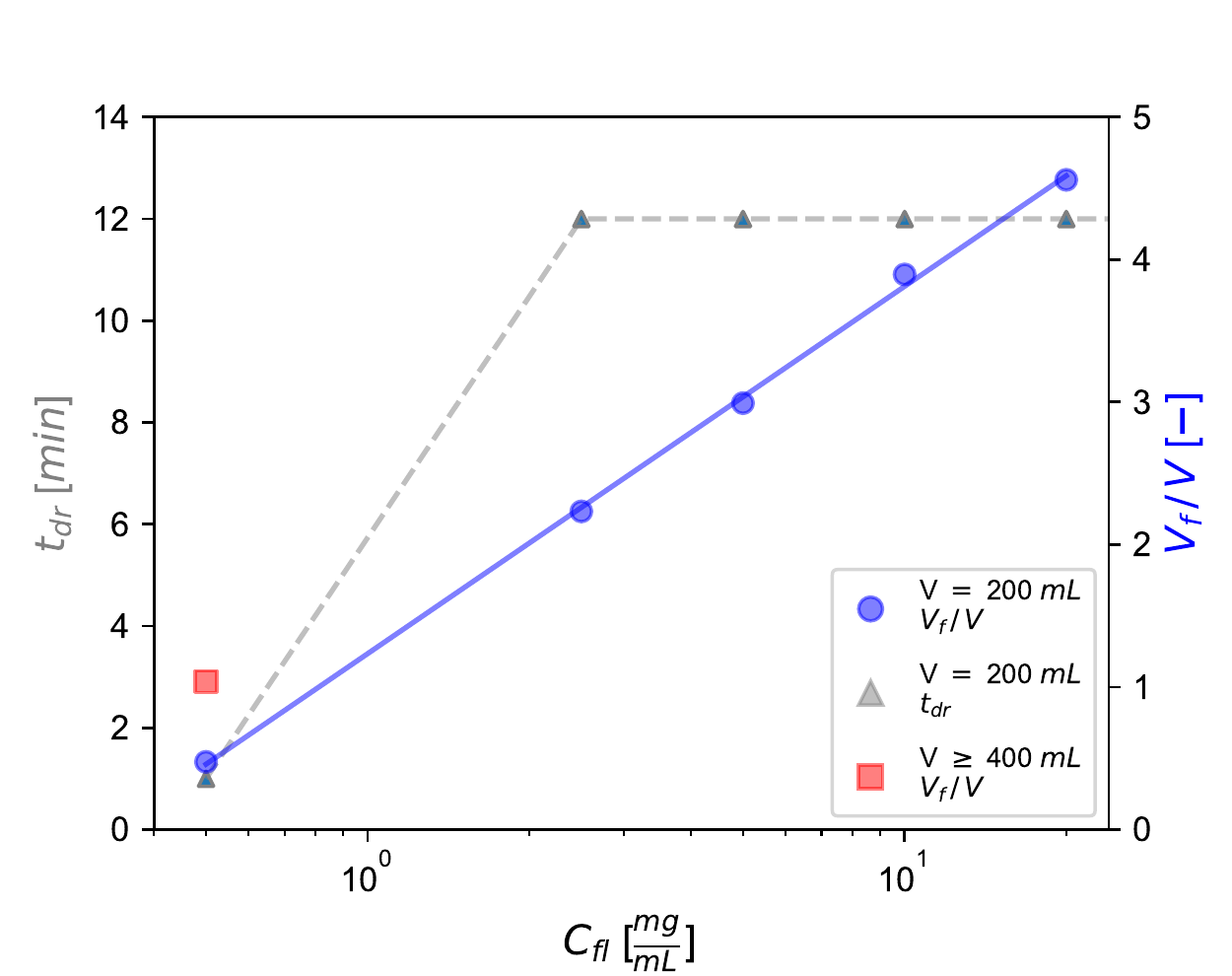}
	\caption{Effect of surfactant concentration on foam volume ($V_f$) and liquid drainage time ($t_{dr}$). Experiments were performed at $\omega = 10000$ rpm, $t = 30$ s and without graphite.}
	\label{fig:10}
\end{figure}

The increase in axial velocity component relative to the circumferential flow in the vessel is a consequence of changes to the foam rheology. As surfactant concentration is increased, the foam volume increases. Figure \ref{fig:10} shows the change in foam volume for different surfactant concentrations measured using ASTM standard procedures \cite{D3519}. Between the lowest and highest surfactant concentrations in Figure \ref{fig:9}, the foam volume increases tenfold. Larger foam viscosities are experienced and this effect increases the natural resistance to flow which supports self-baffling and dampening vortexing due to liquid swirl \cite{Holland1966}. By dampening the vortexing within the vessel, the cirumferential velocity component is reduced and the flow behaviour in this high shear region changes significantly.    

In the upper half of the vessel, the graphite precursor that has been propelled outwards and upwards by the impeller returns due to gravity along pathways that are also influenced by the surfactant concentration. The downward flow angle increases from $\beta = 16^{\circ}$ for the highest surfactant concentration $C_i/C_{fl} = 1$, to $\beta = 43^{\circ}$ for the lowest $C_i/C_{fl} = 40$. Smaller flow angles are observed at higher surfactant concentrations as the larger fluid viscosity reduces the axial component of velocity.

The meeting of the upward and downward graphite particle flows produces different time-average flow patterns within the vessel. For $C_i/C_{fl} \leqslant 4$, saddle points (1) and nodes (2) were observed (Fig. \ref{fig:9}a-d). In locations where the axial velocity component approaches zero, preferential clustering of graphite particles occurred (Fig. \ref{fig:9}c). These particles have been restricted from motion in the circumferential direction due to the physical presence of the baffles. The combination of low axial velocity, high viscosity and viscoplasticity of the foams formed at high surfactant concentration also support this particle hold-up effect. The preferential clustering of graphite particles in this hold-up region is unfavourable for graphene production as the material is unlikely to be exposed to exfoliation conditions when kept away from the rapidly rotating impeller.

For $C_i/C_{fl} \geqslant 8$, liquid swirl becomes more dominant and the topology of the flow pattern changes. Singular points in the flow field of the measurement region are non-existent (Fig. \ref{fig:9}e,f). In addition to the expected change to the fluid stress fields, graphite hold-up was no longer apparent for foam flows at these surfactant concentrations.   

Based on these observations, it is hypothesised that foam rheology is the leading reason for the variation in optimum surfactant concentration. Although we do not have a measure of viscosity for the foams, we have directly measured what the change in rheology does to the graphite precursor material flows. The addition of surfactant has a coupled effect on turbulent shear exfoliation conditions inside the vessel. Different graphite flow behaviours have been observed which alter the turbulent shear mixing conditions at the macroscale. Changes in mixture rheology also influence the stress fields through viscosity. The shear-rate dependent viscosity of foams can be described by a Bingham plastic model behaviour \cite{Kraynik1988}, $\mu(\dot{\gamma},C_{fl}) = \tau_y/\dot{\gamma} + k\lvert\dot{\gamma}\rvert^{n-1}$. Foam viscosity depends on both the shear rate and the surfactant concentration, the yield stress ($\tau_y$), and the flow and consistency indices ($k,n$). Therefore, recalling $P/V = \mu\dot{\gamma}^{2}$ and $C_g \sim \dot{\gamma}t_m$, the concentration of graphene produced is not only sensitive to surfactant repulsion mechanisms at the nanoscale but also the macroscale hydrodynamics. 

Figure \ref{fig:10} shows that larger foam volumes are generated with increasing surfactant concentration. This increases the foam viscosity as demonstrated by the observations of swirl and vortex suppression in the vessel. In Figure \ref{fig:8}, there is a trend of decreasing optimum surfactant concentration with increasing process volume. Approximately twice the surfactant concentration is required to maximise graphene concentration in a 200 mL process volume compared to the 800 mL process volume. The reason for this trend is believed to be related to the differences in foam formation at small and large process volumes. 

The amount of foam generated per unit volume of water ($V_f/V$) for processes with $V \geqslant 400$ mL is double that which is formed at $V = 200$ mL for the same surfactant concentration (Fig. \ref{fig:10}). In the smaller volume process, the impeller is located just below the liquid interface on start-up. The aeration process is less effective at producing bubbly flows throughout than the larger process volumes (e.g. Fig. \ref{fig:3}). Less liquid-gas interface area per unit volume is available for the accumulation of surface-active agents. As a result, the larger volumes require $\approx 50\%$ less surfactant to generate the same foam levels as $V = 200$ mL during turbulent exfoliation. Although this adds another variable to consider for surfactant selection, it also presents an opportunity to design batch processes for low levels of surfactant usage. An advantage of this flexibility for graphene and other 2D materials is that it could reduce post-production washing and annealing requirements that are needed to remove residual surfactant from the product and recover material properties \cite{Lotya2009}. 

\subsection{Inline production monitoring} \label{realtime}
\begin{figure}
	\centering
		\includegraphics[scale=0.6]{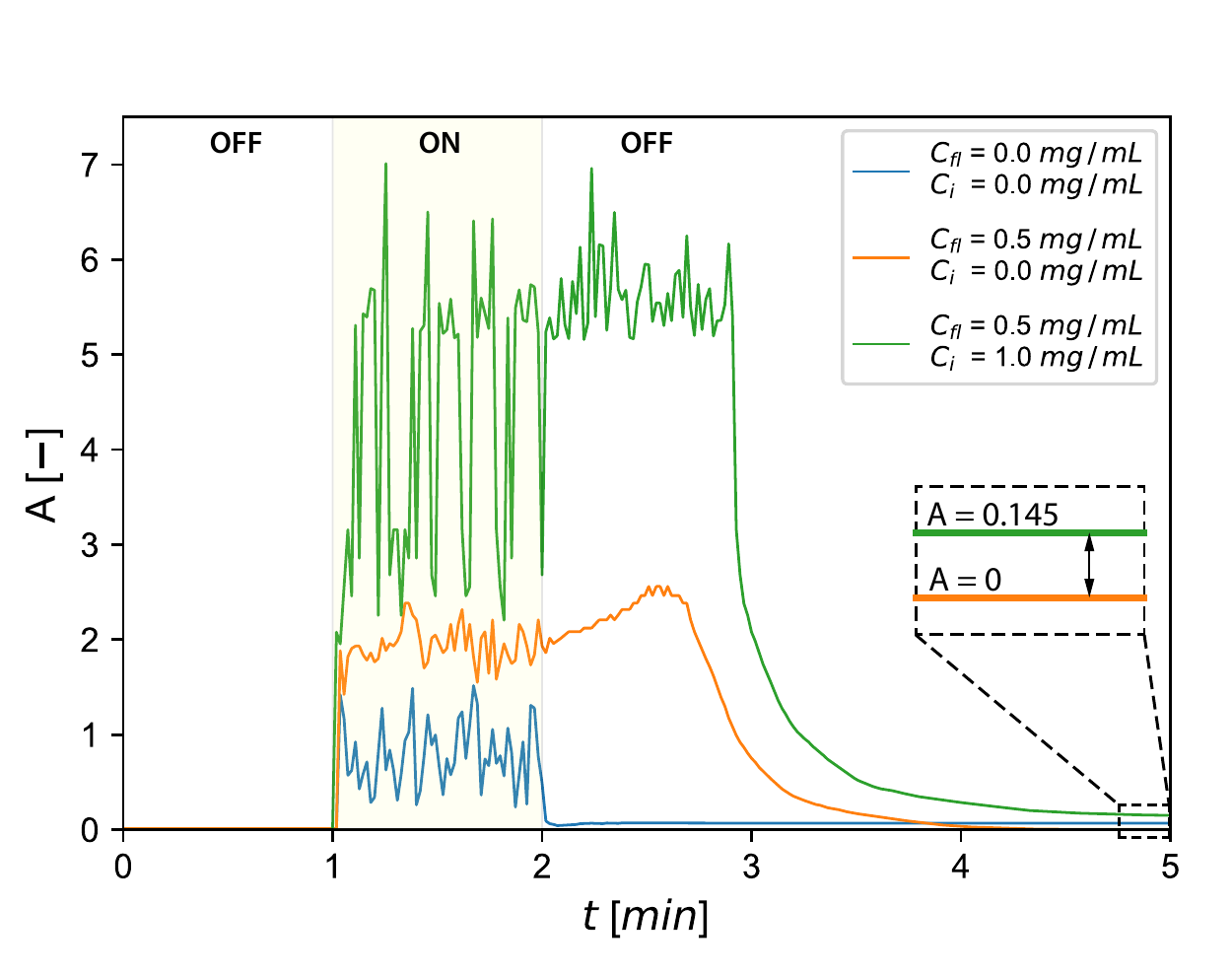}
	\caption{\textit{In situ} spectroscopy measurements of extinction at $\lambda = 660$ nm before, during and after turbulent mixing of water, water-surfactant, and water-surfactant-graphite dispersions. Turbulent mixing occurred between $1\leqslant t\leqslant2$ minutes. Experiments were performed at $\omega = 10000$ rpm and $V = 200$ mL.}
	\label{fig:11}
\end{figure}

\noindent The multi-phase flows in turbulent shear exfoliation of layered materials in aqueous-surfactant dispersions pose a challenge for inline material characterisation using uv-vis-nIR spectroscopy. In section \ref{Insitu-setup}, a provisional assessment and validation of the fibre optic probe system was carried out using water and blue dye. Bubbly flows within the vessel during turbulent mixing at $Re \approx 2.6 \times 10^6$ produced large fluctuations in the optical signal that required sampling to be performed during OFF times in the process. Using water, or water and blue dye, the gas bubbles coalesce within seconds of stopping the mixing process and sampling can begin almost immediately. This behaviour is shown in Figure \ref{fig:11}. 

The addition of surfactants, and the subsequent formation of foams, requires a change in the sampling interval time as bubble coalescence is suppressed. We found this to depend on the liquid drainage time ($t_{dr}$) which describes the time it takes for liquid to drain out of the foam structure when the process is stopped. In the current study, the $t_{dr}$ was determined based on the liquid returning to 85\% of the original fluid height in the vessel, $h_L(0)$ (Fig. \ref{fig:1}c). Figure \ref{fig:10} shows the liquid drainage time for different surfactant concentrations. For $C_{fl} > 1$ mg/mL, the drainage time went beyond 10 minutes. In order to limit the sampling interval time, therefore, inline spectroscopy was performed on dispersions containing $C_{fl} \leqslant 1$.

For water-surfactant mixtures, the foam covers the probe for $\approx 40-50$ s after the impeller is turned OFF (Fig. \ref{fig:11}). Without graphite ($C_i = 0$ mg/mL), the extinction measurements return to the baseline after 2 mins ($A = 0$). When graphite particles are added ($C_i = 1$ mg/mL) and exfoliation begins (ON), the average and fluctuations in extinction increase substantially as both graphitic material and gas bubbles flow across the sensor measurement zone. The addition of the layered precursor resulted in a slightly longer period of time for liquid drainage ($\approx 60$ s). This was followed by a stable decay in extinction, reaching a constant value within 3 minutes of stopping the exfoliation process. The difference in extinction between the baseline and this settled value shown in the inset plot in Figure \ref{fig:11} is a result of exfoliated graphitic material in dispersion. 

\begin{figure*}
	\centering
		\includegraphics[scale=0.65]{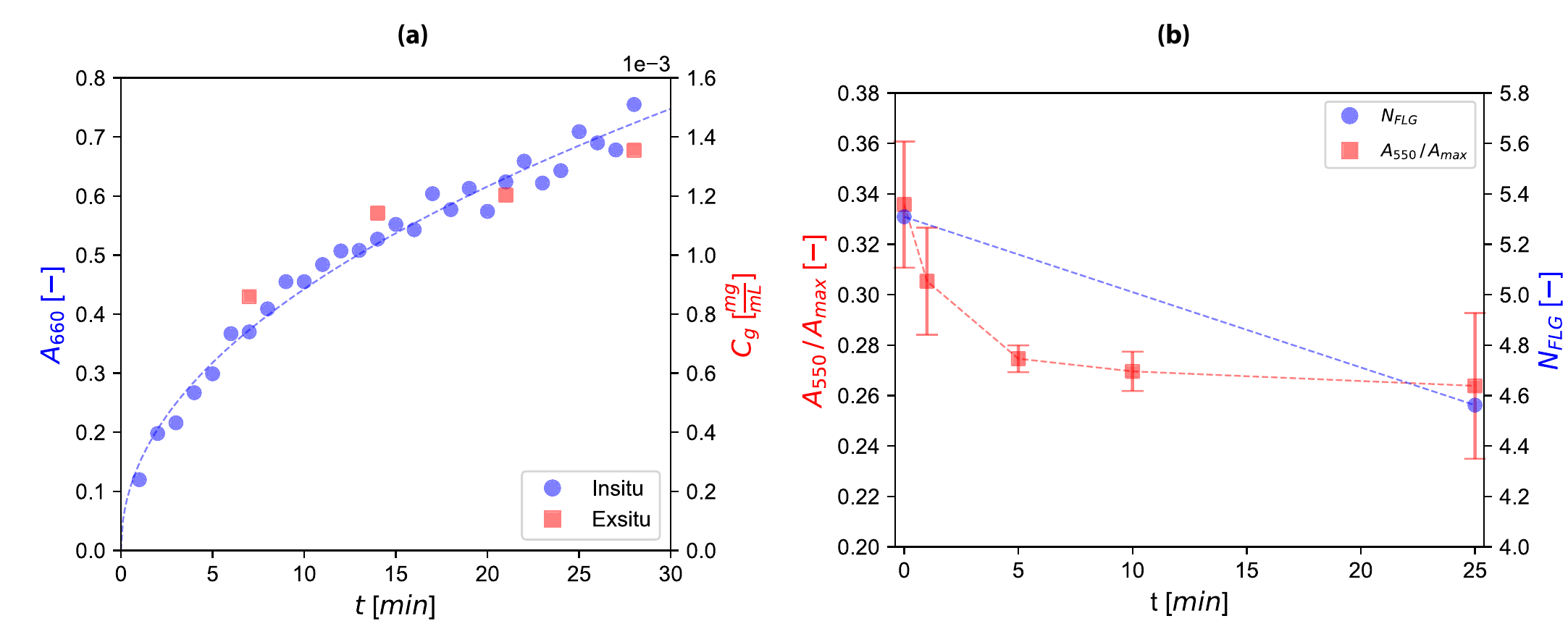}
	\caption{(a) Comparison between \textit{in situ} extinction measurements at $\lambda = 660$ nm and graphene concentration with time. Experiments were performed at $\omega = 10000$ rpm, $V = 200$ mL, $C_i = 1$ mg/mL and $C_i / C_{fl} = 2$. (b) Reduction in average number of atomic layers ($N_{FLG}$) for a few-layer graphene dispersion undergoing turbulent exfoliation. \textit{In situ} measurements of the extinction ratio ($A_{550}/A_{\mathrm{max}}$) corresponding to changes in few-layer graphene spectra with time. Experiments were performed at $\omega = 10000$ rpm, $V = 200$ mL, $C_g = 2.61 \times 10^{-3}$ mg/mL and $C_{fl} \approx 1$ mg/mL.}
	\label{fig:12}
\end{figure*}

Exfoliation was performed for $t = 30$ mins with inline characterisation to investigate if the changes in extinction discussed above correlate with changes in concentration of few-layer graphene. The protocol for performing liquid exfoliation was altered to 1 min (ON) followed by 3 mins (OFF) to permit \textit{in situ} measurements within the heterogeneous dispersions. A reasonable agreement was found between the inline monitoring result and the graphene concentration measurements obtained from physical sampling and $ex situ$ characterisation as shown in Figure \ref{fig:12}a. This suggests the inline characterisation approach and sampling procedures outlined in this work could provide a useful means to estimate production rate in batch exfoliation systems where multi-phase flows exist. These flows preclude real-time information on graphene content using this inline approach. The foam drainage time requirements for monitoring also increase the overall process time. However, the high sampling frequency shown here (1/60 Hz for ON process time) may not be necessary for industrial production systems. For example, four inline measurements required 12 mins whereas four \textit{ex situ} measurements took $\approx 120$ mins to prepare and analyse.         

High-throughput batch quality control is another important area for process scale-up. The properties of graphene and other 2D materials are intrinsically linked with the nanosheet thickness or average number of atomic layers, $N_{FLG}$. To explore if inline spectroscopy could detect changes in layer number, a few-layer graphene dispersion that had been produced in the kitchen blender and centrifuged to remove graphitic material was re-introduced. The process was operated above the critical exfoliation criterion, $\dot{\gamma} > \dot{\gamma}_c$, with the aim of reducing $N_{FLG}$ over time. The extinction spectra was measured before and after the process using \textit{ex situ} uv-vis-nIR spectroscopy. Applying spectroscopic metrics from the work of Backes et al. \cite{Backes2016}, the average layer number was found to reduce by $\Delta{N}_{FLG} = -0.8$ over the 25 minute process time (Fig. \ref{fig:12}b). In parallel, \textit{in situ} spectroscopy measurements were acquired following the same 1 min ON / 3 min OFF procedure previously described. Changes in the spectroscopic metric ($A_{550}/A_{\mathrm{max}}$) that correlate with a reduction in average layer number over time were also observed. Interestingly, this shows that there is a slower rate of change after $t = 10$ mins and highlights an advantage of performing \textit{in situ} monitoring as exfoliation proceeds. Depending on the material application and quality control requirements, the batch exfoliation process could be optimised for tuning layer number while also minimising energy use by avoiding prolonged process times.           

\section{Conclusions}
\noindent Foams can form during turbulent liquid exfoliation of layered materials in aqueous-surfactant dispersions. The impact of foam flows on graphene production and inline spectroscopy has been explored using a kitchen blender to perform batch exfoliation. The optimal surfactant concentration that produced the highest graphene concentration was found to increase with decreasing process volume. Particle image velocimetry measurements on graphite precursor particles revealed that this can be attributed to the hydrodynamics of the batch exfoliation process and the mixture rheology. Foam volume and rheology changes with surfactant concentration and process volume, altering the macroscale flow behaviour that is central to the production of graphene when using turbulent shear exfoliation. The presence of stable foams, persistent gas bubbles and precursor particles also complicates inline 2D material characterisation during exfoliation in aqueous-surfactant dispersions. A measurement protocol has been developed for obtaining concentration and average layer number information during the exfoliation process. Combined with the insights on the hydrodynamics of batch exfoliation in aqueous-surfactants, the findings support the advancement of resource-efficient batch processes with embedded material quality control.         

\section*{Acknowledgements}
\noindent The authors acknowledge funding from The Royal Society (Ref: RGS$\backslash$R2$\backslash$202379) and the British Council Newton Fund - Institutional Links (Ref: 525463513).

\bibliographystyle{elsarticle-num}

\bibliography{cas-refs}

\begin{thebibliography}{10}
\expandafter\ifx\csname url\endcsname\relax
  \def\url#1{\texttt{#1}}\fi
\expandafter\ifx\csname urlprefix\endcsname\relax\def\urlprefix{URL }\fi
\expandafter\ifx\csname href\endcsname\relax
  \def\href#1#2{#2} \def\path#1{#1}\fi

\bibitem{Geim2013}
A.~K. Geim, I.~V. Grigorieva, {Van der Waals heterostructures}, Nature
  499~(7459) (2013) 419--425.
\newblock \href {https://doi.org/10.1038/nature12385}
  {\path{doi:10.1038/nature12385}}.

\bibitem{Nicolosi2013}
V.~Nicolosi, M.~Chhowalla, M.~G. Kanatzidis, M.~S. Strano, J.~N. Coleman,
  Liquid exfoliation of layered materials, Science 340~(6139) (2013).
\newblock \href {https://doi.org/10.1126/science.1226419}
  {\path{doi:10.1126/science.1226419}}.

\bibitem{Ferrari2015}
A.~C. Ferrari, F.~Bonaccorso, V.~Fal{'}ko, K.~S. Novoselov, S.~Roche,
  P.~Bøggild, S.~Borini, F.~H.~L. Koppens, V.~Palermo, N.~Pugno, J.~A.
  Garrido, R.~Sordan, A.~Bianco, L.~Ballerini, M.~Prato, E.~Lidorikis,
  J.~Kivioja, C.~Marinelli, T.~Ryhänen, A.~Morpurgo, J.~N. Coleman,
  V.~Nicolosi, L.~Colombo, A.~Fert, M.~Garcia-Hernandez, A.~Bachtold, G.~F.
  Schneider, F.~Guinea, C.~Dekker, M.~Barbone, Z.~Sun, C.~Galiotis, A.~N.
  Grigorenko, G.~Konstantatos, A.~Kis, M.~Katsnelson, L.~Vandersypen,
  A.~Loiseau, V.~Morandi, D.~Neumaier, E.~Treossi, V.~Pellegrini, M.~Polini,
  A.~Tredicucci, G.~M. Williams, B.~Hee~Hong, J.-H. Ahn, J.~Min~Kim, H.~Zirath,
  B.~J. van Wees, H.~van~der Zant, L.~Occhipinti, A.~Di~Matteo, I.~A. Kinloch,
  T.~Seyller, E.~Quesnel, X.~Feng, K.~Teo, N.~Rupesinghe, P.~Hakonen, S.~R.~T.
  Neil, Q.~Tannock, T.~Löfwander, J.~Kinaret, Science and technology roadmap
  for graphene{,} related two-dimensional crystals{,} and hybrid systems,
  Nanoscale 7 (2015) 4598--4810.
\newblock \href {https://doi.org/10.1039/C4NR01600A}
  {\path{doi:10.1039/C4NR01600A}}.

\bibitem{Stafford2018}
J.~Stafford, A.~Patapas, N.~Uzo, O.~K. Matar, C.~Petit, Towards scale-up of
  graphene production via nonoxidizing liquid exfoliation methods, AIChE
  Journal 64~(9) (2018) 3246--3276.
\newblock \href {https://doi.org/10.1002/aic.16174}
  {\path{doi:10.1002/aic.16174}}.

\bibitem{Hernandez2008}
Y.~Hernandez, V.~Nicolosi, M.~Lotya, F.~M. Blighe, Z.~Sun, S.~De, I.~T.
  McGovern, B.~Holland, M.~Byrne, Y.~K. Gun'Ko, J.~J. Boland, P.~Niraj,
  G.~Duesberg, S.~Krishnamurthy, R.~Goodhue, J.~Hutchison, V.~Scardaci, A.~C.
  Ferrari, J.~N. Coleman, High-yield production of graphene by liquid-phase
  exfoliation of graphite, Nature Nanotechnology 3~(9) (2008) 563--568.
\newblock \href {https://doi.org/10.1038/nnano.2008.215}
  {\path{doi:10.1038/nnano.2008.215}}.

\bibitem{Paton2014}
K.~R. Paton, E.~Varrla, C.~Backes, R.~J. Smith, U.~Khan, A.~O'Neill, C.~Boland,
  M.~Lotya, O.~M. Istrate, P.~King, T.~Higgins, S.~Barwich, P.~May,
  P.~Puczkarski, I.~Ahmed, M.~Moebius, H.~Pettersson, E.~Long, J.~Coelho, S.~E.
  O'Brien, E.~K. McGuire, B.~M. Sanchez, G.~S. Duesberg, N.~McEvoy, T.~J.
  Pennycook, C.~Downing, A.~Crossley, V.~Nicolosi, J.~N. Coleman, Scalable
  production of large quantities of defect-free few-layer graphene by shear
  exfoliation in liquids, Nature Materials 13~(6) (2014) 624--630.
\newblock \href {https://doi.org/10.1038/nmat3944}
  {\path{doi:10.1038/nmat3944}}.

\bibitem{Paton2017}
K.~R. Paton, J.~Anderson, A.~J. Pollard, T.~Sainsbury, Production of few-layer
  graphene by microfluidization, Materials Research Express 4~(2) (2017)
  025604.
\newblock \href {https://doi.org/10.1088/2053-1591/aa5b24}
  {\path{doi:10.1088/2053-1591/aa5b24}}.

\bibitem{Stafford2021}
J.~Stafford, N.~Uzo, U.~Farooq, S.~Favero, S.~Wang, H.-H. Chen,
  A.~L{\textquotesingle}Hermitte, C.~Petit, O.~Matar, Real-time monitoring and
  hydrodynamic scaling of shear exfoliated graphene, 2D Materials 8 (2021)
  025029(1--17).
\newblock \href {https://doi.org/10.1088/2053-1583/abdf2f}
  {\path{doi:10.1088/2053-1583/abdf2f}}.

\bibitem{Chen2014}
X.~Chen, N.~M. Smith, K.~S. Iyer, C.~L. Raston, Controlling nanomaterial
  synthesis{,} chemical reactions and self assembly in dynamic thin films,
  Chem. Soc. Rev. 43 (2014) 1387--1399.
\newblock \href {https://doi.org/10.1039/C3CS60247H}
  {\path{doi:10.1039/C3CS60247H}}.

\bibitem{Varrla2014}
E.~Varrla, K.~R. Paton, C.~Backes, A.~Harvey, R.~J. Smith, J.~McCauley, J.~N.
  Coleman, Turbulence-assisted shear exfoliation of graphene using household
  detergent and a kitchen blender, Nanoscale 6 (2014) 11810--11819.
\newblock \href {https://doi.org/10.1039/C4NR03560G}
  {\path{doi:10.1039/C4NR03560G}}.

\bibitem{Yi2014}
M.~Yi, Z.~Shen, Kitchen blender for producing high-quality few-layer graphene,
  Carbon 78 (2014) 622--626.
\newblock \href {https://doi.org/10.1016/j.carbon.2014.07.035}
  {\path{doi:10.1016/j.carbon.2014.07.035}}.

\bibitem{Biccai2019}
S.~Biccai, S.~Barwich, D.~Boland, A.~Harvey, D.~Hanlon, N.~McEvoy, J.~N.
  Coleman, Exfoliation of 2d materials by high shear mixing, 2D Materials 6~(1)
  (2019) 015008.
\newblock \href {https://doi.org/10.1088/2053-1583/aae7e3}
  {\path{doi:10.1088/2053-1583/aae7e3}}.

\bibitem{Zhang2012}
J.~Zhang, S.~Xu, W.~Li, High shear mixers: A review of typical applications and
  studies on power draw, flow pattern, energy dissipation and transfer
  properties, Chemical Engineering and Processing: Process Intensification
  57-58 (2012) 25--41.
\newblock \href {https://doi.org/10.1016/j.cep.2012.04.004}
  {\path{doi:10.1016/j.cep.2012.04.004}}.

\bibitem{Bakker2004}
A.~Bakker, L.~Oshinowo, Modelling of turbulence in stirred vessels using large
  eddy simulation, Chemical Engineering Research and Design 82~(9) (2004)
  1169--1178.
\newblock \href {https://doi.org/10.1205/cerd.82.9.1169.44153}
  {\path{doi:10.1205/cerd.82.9.1169.44153}}.

\bibitem{Janiga2019}
G.~Janiga, Large-eddy simulation and 3d proper orthogonal decomposition of the
  hydrodynamics in a stirred tank, Chemical Engineering Science 201 (2019)
  132--144.
\newblock \href {https://doi.org/10.1016/j.ces.2019.01.058}
  {\path{doi:10.1016/j.ces.2019.01.058}}.

\bibitem{Vikash2019}
Vikash, V.~Kumar, Turbulent statistics of flow fields using large eddy
  simulations in batch high shear mixers, Chemical Engineering Research and
  Design 147 (2019) 561--569.
\newblock \href {https://doi.org/10.1016/j.cherd.2019.05.045}
  {\path{doi:10.1016/j.cherd.2019.05.045}}.

\bibitem{Nienow1997}
A.~Nienow, On impeller circulation and mixing effectiveness in the turbulent
  flow regime, Chemical Engineering Science 52~(15) (1997) 2557--2565.
\newblock \href {https://doi.org/10.1016/S0009-2509(97)00072-9}
  {\path{doi:10.1016/S0009-2509(97)00072-9}}.

\bibitem{Sinclair2018}
R.~C. Sinclair, J.~L. Suter, P.~V. Coveney, Graphene–graphene interactions:
  Friction, superlubricity, and exfoliation, Advanced Materials 30~(13) (2018)
  1705791.
\newblock \href {https://doi.org/10.1002/adma.201705791}
  {\path{doi:10.1002/adma.201705791}}.

\bibitem{Salussolia2020}
G.~Salussolia, E.~Barbieri, N.~M. Pugno, L.~Botto, Micromechanics of
  liquid-phase exfoliation of a layered 2d material: A hydrodynamic peeling
  model, Journal of the Mechanics and Physics of Solids 134 (2020) 103764.
\newblock \href {https://doi.org/10.1016/j.jmps.2019.103764}
  {\path{doi:10.1016/j.jmps.2019.103764}}.

\bibitem{Zheling2020}
Z.~Li, R.~J. Young, C.~Backes, W.~Zhao, X.~Zhang, A.~A. Zhukov, E.~Tillotson,
  A.~P. Conlan, F.~Ding, S.~J. Haigh, K.~S. Novoselov, J.~N. Coleman,
  Mechanisms of liquid-phase exfoliation for the production of graphene, ACS
  Nano 14~(9) (2020) 10976--10985.
\newblock \href {https://doi.org/10.1021/acsnano.0c03916}
  {\path{doi:10.1021/acsnano.0c03916}}.

\bibitem{Hernandez2010}
Y.~Hernandez, M.~Lotya, D.~Rickard, S.~D. Bergin, J.~N. Coleman, Measurement of
  multicomponent solubility parameters for graphene facilitates solvent
  discovery, Langmuir 26~(5) (2010) 3208--3213.
\newblock \href {https://doi.org/10.1021/la903188a}
  {\path{doi:10.1021/la903188a}}.

\bibitem{Salavagione2017}
H.~J. Salavagione, J.~Sherwood, M.~De~bruyn, V.~L. Budarin, G.~J. Ellis, J.~H.
  Clark, P.~S. Shuttleworth, Identification of high performance solvents for
  the sustainable processing of graphene, Green Chem. 19 (2017) 2550--2560.
\newblock \href {https://doi.org/10.1039/C7GC00112F}
  {\path{doi:10.1039/C7GC00112F}}.

\bibitem{Lotya2009}
M.~Lotya, Y.~Hernandez, P.~J. King, R.~J. Smith, V.~Nicolosi, L.~S. Karlsson,
  F.~M. Blighe, S.~De, Z.~Wang, I.~T. McGovern, G.~S. Duesberg, J.~N. Coleman,
  Liquid phase production of graphene by exfoliation of graphite in
  surfactant/water solutions, Journal of the American Chemical Society 131~(10)
  (2009) 3611--3620.
\newblock \href {https://doi.org/10.1021/ja807449u}
  {\path{doi:10.1021/ja807449u}}.

\bibitem{Lotya2010}
M.~Lotya, P.~J. King, U.~Khan, S.~De, J.~N. Coleman, High-concentration,
  surfactant-stabilized graphene dispersions, ACS Nano 4~(6) (2010) 3155--3162.
\newblock \href {https://doi.org/10.1021/nn1005304}
  {\path{doi:10.1021/nn1005304}}.

\bibitem{Backes2016}
C.~Backes, K.~R. Paton, D.~Hanlon, S.~Yuan, M.~I. Katsnelson, J.~Houston, R.~J.
  Smith, D.~McCloskey, J.~F. Donegan, J.~N. Coleman, Spectroscopic metrics
  allow in situ measurement of mean size and thickness of liquid-exfoliated
  few-layer graphene nanosheets, Nanoscale 8 (2016) 4311--4323.
\newblock \href {https://doi.org/10.1039/C5NR08047A}
  {\path{doi:10.1039/C5NR08047A}}.

\bibitem{Harnby1992}
N.~Harnby, M.~F. Edwards, A.~W. Nienow, Mixing in the Process Industries, 2nd
  Edition, Butterworth-Heinemann, Oxford, 1992.

\bibitem{Kraynik1988}
A.~M. Kraynik, Foam flows, Annual Review of Fluid Mechanics 20~(1) (1988)
  325--357.
\newblock \href {https://doi.org/10.1146/annurev.fl.20.010188.001545}
  {\path{doi:10.1146/annurev.fl.20.010188.001545}}.

\bibitem{Holland1966}
F.~A. Holland, F.~S. Chapman, Liquid Mixing and Processing in Stirred Tanks,
  Reinhold Publishing Corporation, New York, 1966.

\bibitem{PIVlab}
W.~Thielicke, E.~Stamhuis, {PIVlab – Towards} user-friendly, affordable and
  accurate digital particle image velocimetry in {MATLAB}, Journal of Open
  Research Software 2~(1) (2014) p.e30.
\newblock \href {https://doi.org/10.5334/jors.bl} {\path{doi:10.5334/jors.bl}}.

\bibitem{Stafford2012}
J.~Stafford, E.~Walsh, V.~Egan, A statistical analysis for time-averaged
  turbulent and fluctuating flow fields using particle image velocimetry, Flow
  Measurement and Instrumentation 26 (2012) 1--9.
\newblock \href {https://doi.org/doi.org/10.1016/j.flowmeasinst.2012.04.013}
  {\path{doi:doi.org/10.1016/j.flowmeasinst.2012.04.013}}.

\bibitem{SanchezPerez2006}
J.~{Sánchez Pérez}, E.~{Rodríguez Porcel}, J.~{Casas López}, J.~{Fernández
  Sevilla}, Y.~Chisti, Shear rate in stirred tank and bubble column
  bioreactors, Chemical Engineering Journal 124~(1) (2006) 1--5.
\newblock \href {https://doi.org/https://doi.org/10.1016/j.cej.2006.07.002}
  {\path{doi:https://doi.org/10.1016/j.cej.2006.07.002}}.

\bibitem{Nagata1975}
S.~Nagata, Mixing : {P}rinciples and applications, Kodansha : Wiley, Tokyo ;
  New York, 1975.

\bibitem{D3519}
{Standard Test Method for Foam in Aqueous Media ({Blender Test})}, Standard,
  ASTM International, PA, USA (2007).

\end{thebibliography}





\end{document}